\documentclass[twocolumn]{aastex62}
\pdfoutput=1 
\usepackage{amsmath,amstext}
\usepackage[T1]{fontenc}
\usepackage[figure,figure*]{hypcap}
\graphicspath{{./}{figures/}}

\received{?}
\revised{?}
\accepted{?}
\submitjournal{AAS}

\shorttitle{Larger mutual inclinations for the shortest-period planets}
\shortauthors{Dai et al.}

\begin{document}

\title{Larger mutual inclinations for the shortest-period planets}

\author[0000-0002-8958-0683]{Fei Dai}
\affiliation{Department of Physics and Kavli Institute for Astrophysics and Space Research,\\Massachusetts Institute of Technology, Cambridge, MA, 02139, USA}
\affiliation{Department of Astrophysical Sciences, Princeton University, 4 Ivy Lane, Princeton, NJ, 08544, USA}
\email{fd284@mit.edu}

\author[0000-0003-1298-9699]{Kento Masuda}
\affiliation{Department of Astrophysical Sciences, Princeton University, 4 Ivy Lane, Princeton, NJ, 08544, USA}
\affiliation{NASA Sagan Fellow}

\author[0000-0002-4265-047X]{Joshua N.\ Winn}
\affiliation{Department of Astrophysical Sciences, Princeton University, 4 Ivy Lane, Princeton, NJ, 08544, USA}



\begin{abstract}
  The {\it Kepler} mission revealed a population of compact
  multiple-planet systems with orbital periods shorter than a year,
  and occasionally even shorter than a day.
  By analyzing a sample of 102 {\it Kepler} and {\it K2} multi-planet systems, we
  measure the minimum difference $\Delta I$ between the orbital inclinations, as a
  function of the orbital distance of the innermost planet.  This is
  accomplished by fitting all the planetary
  signals simultaneously, constrained by an external estimate of the
  stellar mean density.  We find $\Delta I$ to be larger when the inner
  orbit is smaller, a trend that does not appear to be a selection effect.
  We find that planets with $a/R_\star$<5 have a
  dispersion in $\Delta I$ of $6.7\pm 0.6$~degrees, while planets with $5 <
  a/R_\star < 12$ have a dispersion of $2.0\pm 0.1$~degrees. The
  planetary pairs with higher mutual inclinations also tend to have
  larger period ratios.  These trends suggest that the shortest-period
  planets have experienced both inclination excitation and orbital
  shrinkage.
\end{abstract}




\section{Introduction} \label{sec:intro}

One of the revelations of the {\it Kepler} mission was that Sun-like
stars often host planets with sizes between those of Earth and
Neptune and orbital periods shorter than a year
\citep{Borucki+2011}.  The formation of these short-period planets and
their relationship to wider-orbiting planets are not understood.  An
interesting clue is that the population of planets with the 
shortest periods ($\lesssim$10 days) is different, in some
respects, from the population with longer periods.

One difference is in the planet occurrence rate.
The function $d\log N/d\log P$, where $N$ is the mean number of
planets per star and $P$ is the orbital period, increases with period
from 0.2--10 days before leveling off to a constant value out to 
at least 100 days \citep{Petigura}.  Another difference is
that stars hosting sub-Neptune planets with periods shorter than about 10 days
tend to have higher metallicities than those hosting planets with
longer periods \citep{Mulders+2016,Petigura,Wilson+2018}. 
A third difference is in the period ratios between adjacent planets.
When the inner planet's period is shorter than
a few days, the period ratio tends to be larger
than when both planets have longer periods \citep{Steffen}.

This {\it Letter} describes another clue, which we found in the distribution
of mutual inclinations.
Several scenarios have been proposed
to explain the shortest-period planets.
In almost all these scenarios, the planet's orbit is initially wider,
because of the theoretical difficulty of building
a rocky core in close proximity to the star.
Some of the proposed mechanisms to shrink the orbits also involve
raising the inclination \citep{Hansen,Petrovich}, while others
predict low inclinations \citep[e.g.,][]{Lee}.
Previous studies of {\it Kepler} systems concluded that
the mutual inclinations are typically $\lesssim$$5^\circ$,
based on population statistics \citep{Tremaine,Fabrycky,FangMargot2012}. Here, we
focus on systems with the closest-orbiting planets, and
attempt to measure the mutual inclination of each system directly
by fitting the transit light curves. 
\clearpage
\section{Sample Selection}\label{sec:sample}

Any sample of planets detected with the transit method is strongly
biased against systems with large mutual inclinations.  However, as
the planet's orbit becomes smaller, the range of transiting inclinations
becomes larger.  The requirement is $\cos I < R_\star/a$, where $I$ is the
inclination, $R_\star$ is the stellar radius, and $a$ is the orbital
distance.  For $a/R_\star=4$ (corresponding to $P\approx 1$~day for a Sun-like star),
transits can occur for inclinations between 75--90$^\circ$.
Therefore, by measuring the inclinations of innermost pair of transiting
planets around the same star, we can place a lower bound on the
mutual inclination ranging up to $15^\circ$ for the most favorable
cases.

For this study, we selected the {\it Kepler}
multiple-planet systems with apparent {\it Kepler} magnitude ($Kp$) brighter
than 14, for which the
innermost planet has a radius smaller than $4\,R_\oplus$,
a transit signal-to-noise ratio (SNR)
greater than 20, and $a/R_\star < 12$.
The limit on $Kp$ and SNR ensures that the stars have been well characterized
and the transit signals can be modeled precisely.  The limit on planet size
excludes giant planets, which may have a completely different history
of formation and evolution than smaller planets.
The limit on $a/R_\star$ corresponds to an allowed range of
inclinations of 85--90$^\circ$.
We augmented our sample with planets with periods $<$1~day discovered with {\it K2} data.
We expect the {\it Kepler} and {\it K2} systems to be drawn from similar
populations, since the observations
were made with the same telescope and achieved nearly equivalent
photometric precision after correcting for {\it K2} systematics.
Table~1 reports the most important characteristics of the sample.

\section{Constraints on mean stellar density}\label{sec:density}

When fitting a transit light curve, there is often a strong covariance
between $I$ and $a/R_\star$.  To reduce
this covariance, we enforced
an external constraint on the stellar mean density $\rho_\star$,
which is related to $a/R_\star$ through Kepler's third law.
The external constraint came from fitting
stellar-evolutionary models to the observed spectroscopic parameters,
$K$-band apparent magnitude, and parallax.  We used
spectroscopic parameters from the California-{\it Kepler} survey
\citep{2017AJ....154..107P} whenever available,
and from other sources as needed (see Table 1).
We imposed a minimum uncertainty of 100\,K in the effective temperature
to account for possible systematic errors.  The $K$ magnitudes were
from the Two Micron All Sky Survey \citep[2MASS;][]{2006AJ....131.1163S}, after
correcting for extinction using $A_K=0.443\,E(B-V)$ with the
value of $E(B-V)$ from \cite{2016ApJ...818..130B} (and adopting a 30\% uncertainty).
The parallaxes were from the {\it Gaia} Data Release 2 \citep{2018arXiv180409365G}, although in practice
we used the distance estimates provided by \cite{2018arXiv180410121B}.
The stellar-evolutionary models were from the Dartmouth Stellar Evolution Database \citep{2008ApJS..178...89D} and
were compared to the data using the {\tt isochrones} package by
\citet{2015ascl.soft03010M}.  The results are
given in Column 2 of Table 1.  The typical uncertainty is 8\%.

As a consistency check, we compared the isochrone-based mean densities
with those that were derived from asteroseismology by
\citet{vaneylen}.  On average, the isochrone densities are 5\% smaller than the asteroseismic densities, with a dispersion of 5\%.  This
suggests that one or both methods are subject to systematic errors of a few percent.
This level of error does not have an
appreciable effect on the subsequent results, as we verified
directly, by repeating the analysis using the asteroseismic densities
in place of the isochrone-based densities whenever both were available.

\section{Light-curve analysis}\label{sec:modeling}

For the {\it Kepler} systems we used the Pre-search Data Conditioning
light curves, and for the {\it K2} systems we used the target pixel
files, both were obtained from the Mikulski Archive for Space
Telescopes\footnote{\url{https://archive.stsci.edu}}. To
construct the {\it K2} time series and mitigate
the systematics from the rolling motion of the spacecraft, we used
the code described by \cite{Dai2017}.  Prior to
transit modeling, we removed any long-term photometric
trends due to stellar variability or instrumental effects.  We masked
out the known transits, and fitted the transit-free light
curve with a cubic spline of width 0.75~days.
In cases for which a nearby stellar companion was reported
by \citet{Furlan}, we corrected the light curve for the ``diluting''
effect of the companion.

For each transit, we isolated the segment of data spanning
three times the transit duration, centered on the midpoint.
We visually inspected each transit and removed those few that were
obviously damaged by systematic effects.
We used the {\tt Batman} code for transit
modeling \citep{Kreidberg2015}.  The free parameters were the orbital
period $P$, the midtransit time ($T_{\rm c}$), the planet-to-star
radius ratio ($R_{\rm p}/R_\star$), the scaled orbital distance
($a/R_\star$), and $\cos I$. We adopted a quadratic
limb darkening profile, with coefficients
subject to Gaussian priors with widths of 0.3 and mean values
determined with {\tt EXOFAST}\footnote{\url{astroutils.astronomy.ohio-state.edu/exofast/limbdark.shtml}.} \citep{Eastman2013}.
For the long-cadence data, we computed
the model light curve at 1-minute intervals and averaged it to 30~minutes
before comparing it to the data.

To account for possible transit-timing variations,
we used an iterative process.
First, a constant-period transit model was optimized based on all
the data.  Next, this model
was used as a template to derive individual transit times,
by refitting each transit with the free parameters
limited to the midtransit time and a quadratic
function of time to account for stellar variability.
Then, we fitted a linear function of epoch to the
transit times, to see if this fit was satisfactory, or if a sinusoidal
model provided a better description. We combined all the data to
create a phase-folded light curve,
where the folding was based on either a constant period, or the
individually measured transit times if TTVs had been detected.  This
phase-folded light curve was then fitted, leading to an improved
template for measuring individual transit times. This process
converged after 2-3 iterations. We did not end up identifying
any TTVs that had not already been flagged by \cite{Ofir}.

Finally, we fitted the phase-folded light
curves for all the planets in each system simultaneously, while imposing an
external constraint on the stellar mean density (Section~\ref{sec:density}).
This was done with the
Markov Chain Monte Carlo method as implemented by \cite{emcee}.
We imposed a Jeffreys prior on $R_{\rm p}/R_\star$ and a uniform prior
on $\cos I$.   We restricted $\cos I>0$
because transit data alone cannot be used to determine
the sign of $\cos I$. We assumed the orbit of the innermost planet to be circular, since
tidal circularization is expected to be rapid for $a/R_\star <12$.
The outer planets were allowed to have eccentric orbits.
Uniform priors were adopted for
$\sqrt{e}\cos\omega$ and $\sqrt{e}\sin\omega$, where $e$ is the
eccentricity and $\omega$ is the argument of pericenter.

\begin{figure*}
\begin{center}
\includegraphics[width = .67\columnwidth]{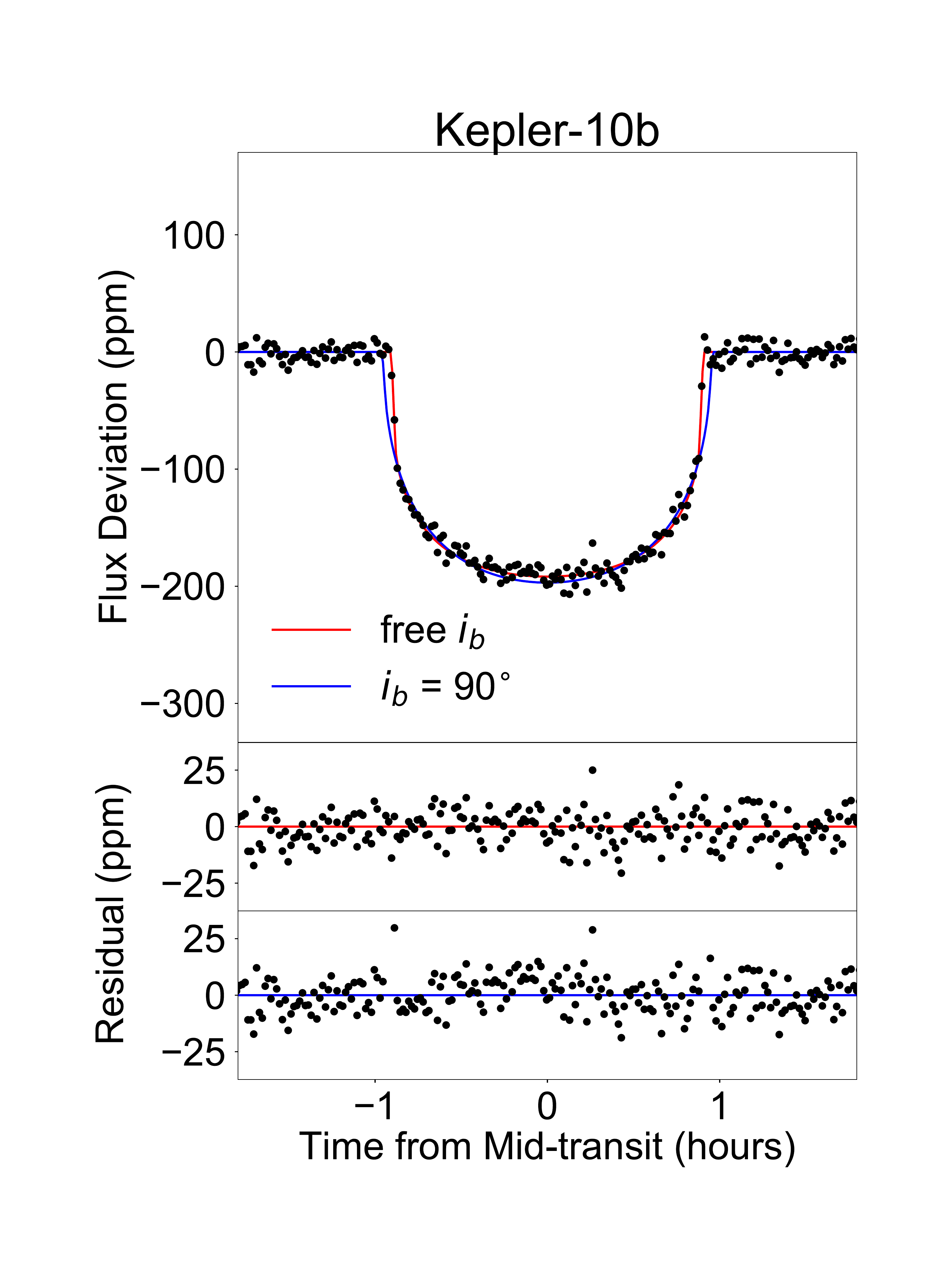}
\includegraphics[width = .67\columnwidth]{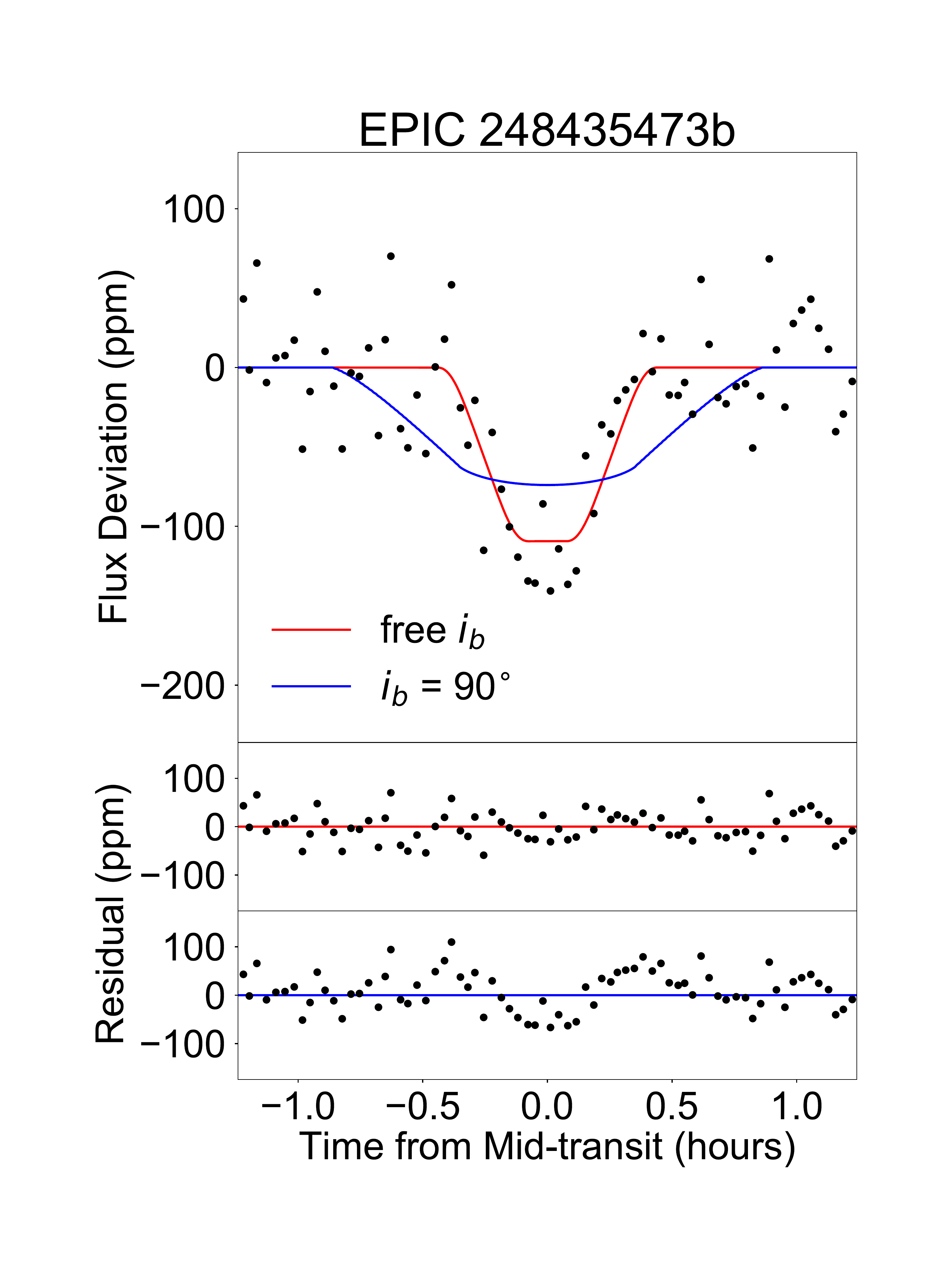}
\includegraphics[width = .67\columnwidth]{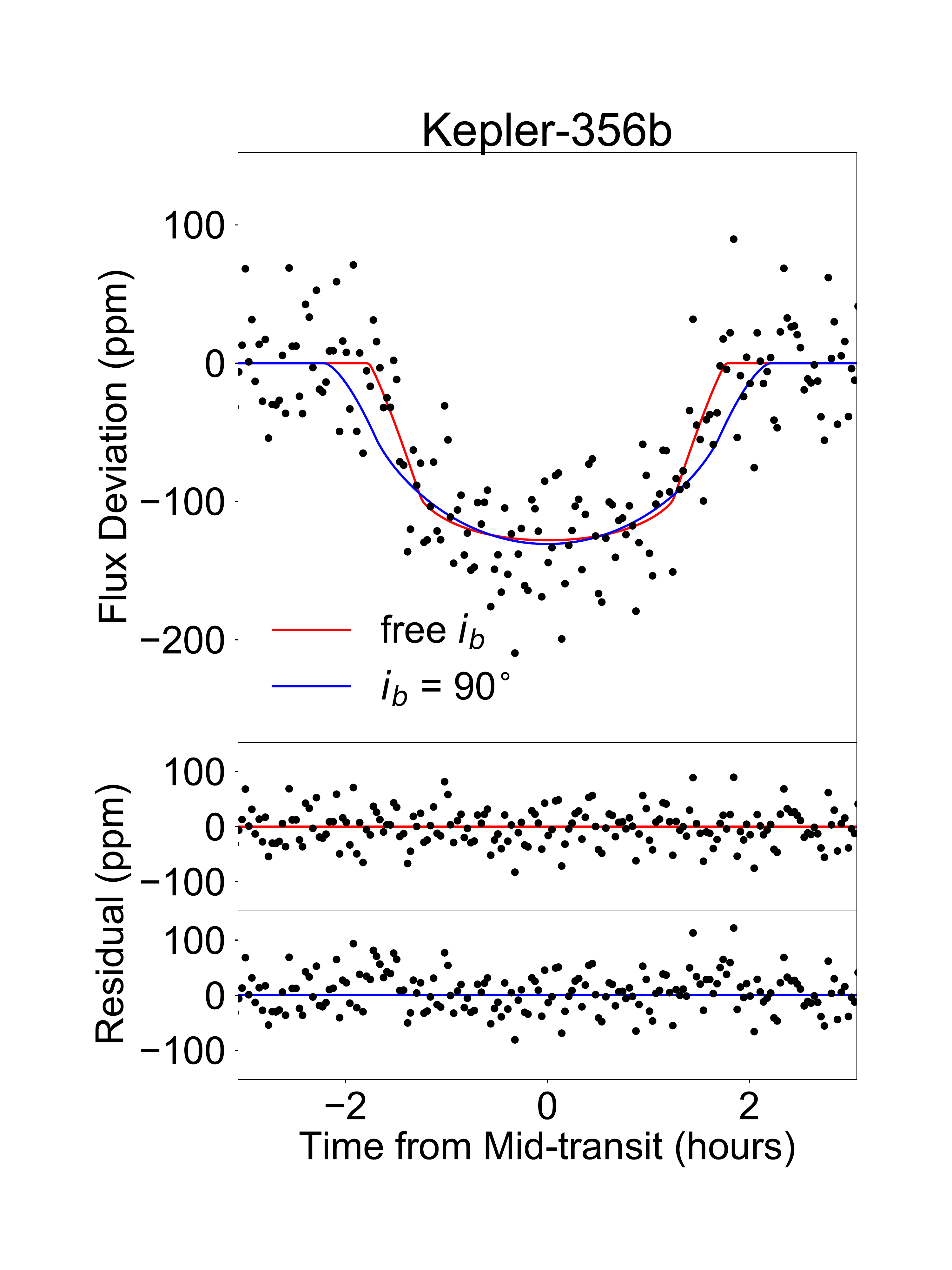}

\caption{Signatures of inclination in the light curves.  Shown are
  three representative phase-folded light curves, along with the
  best-fitting model (red curve), the best fitting model with
  $I=90^\circ$ (blue curve), and the corresponding residuals.  In all
  these cases, there are patterned residuals in lowest panel,
  indicating that $I=90^\circ$ is disfavored.}
\label{fig:lc}
\end{center}
\end{figure*}

Table 1 reports the results for $a/R_\star$ and $I$.
For a visual appreciation of the task
of measuring inclinations, Figure~\ref{fig:lc} shows three
representative light curves.  Also shown are the residuals between the
data and two different models: one in which the inclination is allowed
to vary freely, and one in which it is held fixed at $90^\circ$.

\begin{figure*}
\begin{center}
\includegraphics[width = 2.\columnwidth]{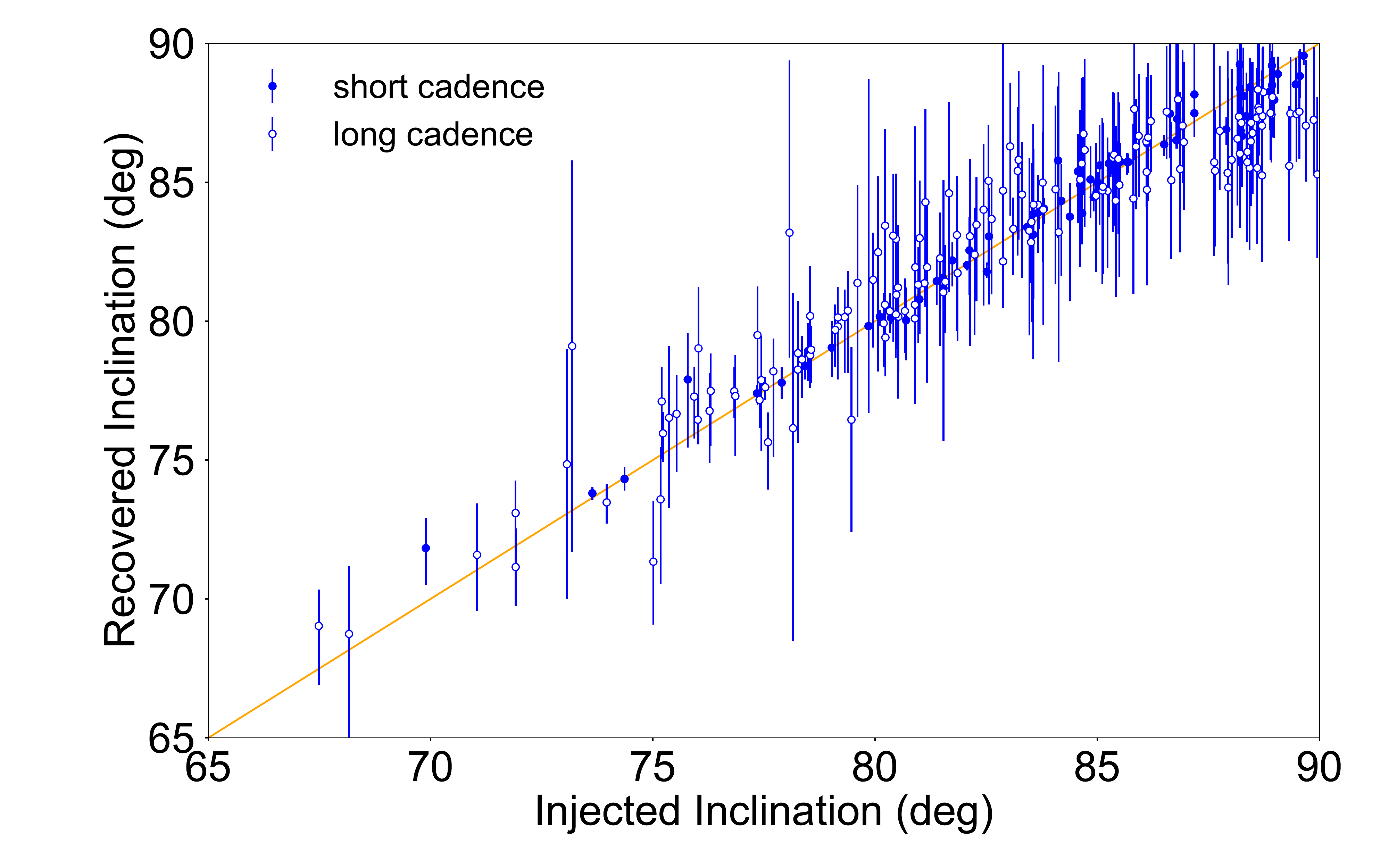}
\caption{Results of the inject-and-recover test for orbital inclination.
  Filled circles are for systems observed in short-cadence mode (1-minute sampling)
  and open circles are for long-cadence mode (30-minute).}
\label{fig:injection}
\end{center}
\end{figure*}

As a test of 
robustness, we repeated the entire analysis,
allowing eccentric orbits for all planets.  Naturally,
this broadened the allowed range of inclinations, but led to no
qualitative changes in the results.
For further validation,
we performed inject-and-recovery simulations.  
We focused on the planets with periods shorter than one day, for which
the coarse time sampling is of greatest concern.  Beginning with the original
time series (prior to any detrending), we injected a transit signal
into the data with the parameters of the best-fitting model but
a different midtransit time and orbital inclination.
The fake signals were subjected to the same
procedures as the real signals.  This was repeated for 5 choices of
$\cos I$ over the range for which transits occur.
Figure~\ref{fig:injection} compares the injected and recovered inclinations,
with error bars based on the 68\% credible
intervals.  The agreement is almost
always within 1-$\sigma$, lending confidence to the results.  Toward
$90^\circ$, the recovered inclination always falls below the identity line, but
this is only because we chose to plot the median of the
posterior, which is necessarily lower than $90^\circ$.

\section{Minimum Mutual Inclinations}\label{sec:results}

For each system, we computed $\Delta I =
|I_1-I_2|$, where $I_1$ and $I_2$ are the fitted orbital inclinations.
In general, $\Delta I$ is a lower bound on the
mutual inclination.  It is equal to the mutual
inclination only if the two planets' trajectories across the stellar
disk are parallel and on the same hemisphere of the star.

\begin{figure*}
\begin{center}
\includegraphics[width = 2.2\columnwidth]{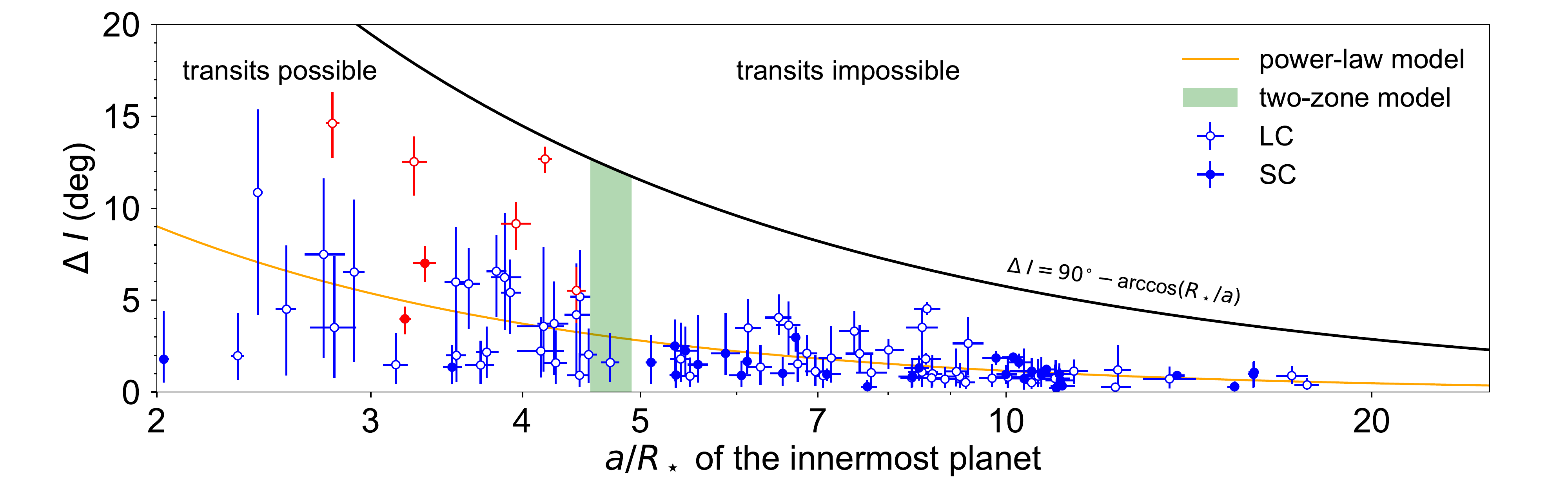}
\includegraphics[width = 2.2\columnwidth]{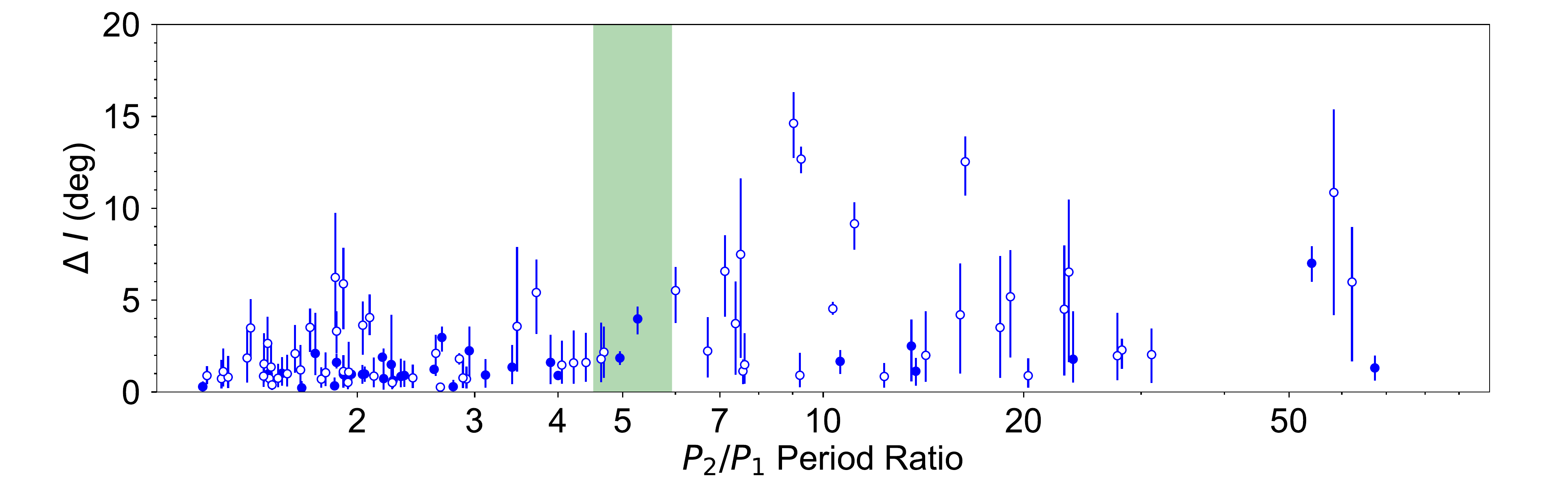}
\caption{Upper Panel: inclination difference $\Delta I$ versus $a/R_\star$ of the innermost
planet.  Filled points are for systems observed in short-cadence mode and hollow points are for long-cadence.  The black line is the boundary above which the inner
planet would not transit, assuming $i_2=90^\circ$.  The orange line
represents Model 2.  The green zone represents the 68\% credible
region for the critical value of $a/R_\star$, in Model 3.  Highlighted
in red are those data points that are more than 3 standard deviations away from
zero: Kepler-10, EPIC~248435473, K2-223, Kepler-312, WASP-47,
KOI-2393 and Kepler-653. Lower Panel:  $\Delta I$ versus orbital period
  ratio $P_2/P_1$.  The green zone is the
  68\% credible interval for the critical value of $P_2/P_1$, in
  Model 3.}
\label{fig:a_r}
\end{center}
\end{figure*}

Figure~\ref{fig:a_r} shows $\Delta I$ as a function of $a/R_\star$ for the innermost planet.
Among the systems with the closest-orbiting planets, there are about
10 systems for which $\Delta I > 5^\circ$, larger than the typical
mutual inclinations that have been inferred for wider-orbiting
planets.\footnote{The relatively large mutual inclination of EPIC~248435473b
($\Delta I = 12.67^{+0.68}_{-0.75}$~deg)
was also noted by \cite{Rodriguez2018}.}
The planets with $a/R_\star >5$ almost all have
$\Delta I$ < 5$^{\circ}$, even though values of
5--10$^\circ$ could have been detected in many cases.
This is consistent with previous studies of the {\it Kepler} multi-planet
systems, which concluded that the mutual inclinations are likely to be
a few degrees or smaller \citep{Tremaine,Fabrycky,FangMargot2012}.
The planets with smaller values of
$a/R_\star$ have a broader distribution of $\Delta I$, nearly filling
the full range of inclinations compatible with transits.

Figure~\ref{fig:a_r} also shows $\Delta I$ as a function of the
period ratio between the innermost two planets. Higher values
of $\Delta I$ are associated with larger period ratios.  This
reflects a pattern noted by \citet{Steffen}: the period
ratios tend to be higher when the inner planet's period
is shorter than about one day.  In that sense, the innermost planets
are dynamically more separated.  Our results show
that these planets are also associated with minimum mutual
inclinations ranging up to 10$^\circ$, higher than that of the broader population of
multi-planet systems.

We quantified these impressions with a hierarchical Bayesian
analysis. We compared the Bayesian evidence for the following models
for the $\Delta I$ distribution:
\begin{enumerate}
  \item A Rayleigh distribution, with width $\sigma$.
  \item A Rayleigh distribution in which the width varies with
    orbital distance: $\sigma = \sigma_0~(a/R_\star)^m$.
  \item A Rayleigh distribution in which the width changes
    abruptly from $\sigma_{\rm in}$ to $\sigma_{\rm out}$ at a critical
    value of $a/R_\star$.
  \item A Rayleigh distribution in which the width changes
    abruptly from $\sigma_{\rm in}$ to $\sigma_{\rm out}$ at a critical
    value of the period ratio, $P_2/P_1$.
\end{enumerate}
To compute the likelihood as a function of the
model parameters, we followed the procedure of
\citet{Foreman-Mackey}, using the approximation
\begin{equation}
 p (\mathrm{obs} | \theta) \propto \prod_{k=1}^{K} \frac{1}{N} \sum_{n=1}^{N} \frac{p(\Delta I_k^{n} | \theta)}{p_0(\Delta I_k^n)},
\end{equation}
where ``obs'' represents the data,
$k$ specifies the planetary system, $n$ specifies random
samples from the posterior of $\Delta I$, and $\theta$ is the set of
hyperparameters.
The function $p_0$
is the prior probability in our transit modeling (Section \ref{sec:modeling}).
Since we adopted uniform and independent priors for $\cos I_1$ and $\cos I_2$, it
can be shown that 
\begin{equation}
p_0(\Delta I) = \frac{1}{16}(\pi - 2\Delta I)\cos(\Delta I).
\end{equation}

All the hyperparameters were subject to log-uniform priors, except for the
exponent $m$ in Model 2 for which we used a uniform prior.  For each
model, we determined the credible intervals for the hyperparameters
and the Bayesian evidence $Z$ using the nested sampling code {\sc
  MultiNest} \citep{Feroz2009}.  Table 2 gives the results.  Models 2
and 3 are both favored over Model 1 by
$\Delta\log Z > 26$, corroborating the visual impression that $\Delta I$ is larger for smaller orbits.
Model 4 is not as successful as Models 2 and 3 but still favored over Model 1 by $\Delta\log Z > 14$.

\section{Discussion}\label{sec:discussion}
We found that when the innermost planet has $a/R_\star \lesssim 5$
(or $P= 1.3$ days for a Sun-like star), the minimum mutual inclination
is often 5--10$^\circ$.
This is somewhat higher than the typical value of a few degrees that has been
previously estimated for the more general population of {\it Kepler} systems.
We also found $\Delta I$ to decrease with the orbital separation of the innermost
planet.  This observation does not appear to be purely a selection effect because
for planets with $a/R_\star$ between 5 and 10, we could have detected mutual inclinations
larger than 5$^\circ$, and we did not.

These results may be related to some previously noted trends.
\citet{Steffen2016} found that the {\it Kepler} sample of
``hot Earths'' without additional transiting companions
is larger than what one would obtain by drawing planets randomly from
the multiple-transiting systems. A related observation by
\citet{Weiss2018} is that the fraction of {\it Kepler} systems with
multiple transiting planets is lower when the innermost planet
has a period shorter than a few days.
Our results offer a natural explanation: the shortest-period planets tend to have larger mutual inclinations, and thus, are more likely to be observed to transit
even when the wider-orbiting companions do not transit.
\citet{Zhu} also found evidence for relatively high mutual inclinations
in some {\it Kepler} systems, based on the observed frequencies of multiple
transiting planets and TTVs. In their model, the inclination dispersion
depends on the total number of planets in the system, ranging from $0.8^\circ$ in five-planet systems to $\sim$10$^\circ$ for two-planet systems.
It would be interesting to try and extend the model of \citet{Zhu} to allow
the mutual inclination to depend on orbital separation in addition to, or instead
of, the total number of planets. \citet{2017MNRAS.468.1493B} proposed that the inner protoplanetary disk may have a flat (rather than flared) geometry in which case the innermost planets tend to form with larger mutual inclinations.

We emphasize that $\Delta I$ only represents a lower bound on the mutual inclination.
Moreover, if there exist systems with much larger mutual inclinations, they
would be unlikely to appear in our sample, because the joint
transit probability is low.  Despite these limitations,
our results indicate that the shortest-period planets have
a different orbital architecture, with higher mutual inclinations and larger period ratios.
This suggests that whatever processes led to the extremely tight orbits of these planets
were also responsible for tilting the orbit to higher inclination.

Several theories have been offered for the formation of very
short-period sup-Neptune planets, which differ in their predictions
for mutual inclinations. \citet{Lee} proposed that the magnetospheric
truncation radius determines the innermost orbit where planets can form.
In this scenario, planets begin with nearly-circular, well-aligned
orbits, and the innermost planet undergoes tidal orbital decay.There is no obvious
agent for exciting inclinations, and therefore this scenario
does not provide an explanation for the larger mutual inclinations of the
shortest-period planets. Likewise, the formation scenarios proposed by \citet{Terquem} and \citet{Schlaufman} do not provide an obvious way to excite inclinations.

\citet{Spalding2016} proposed a scenario that does involve inclination
excitation.
If the host star were initially rotating rapidly, with a non-zero obliquity,
the planets' orbits would undergo nodal precession at different rates
and become misaligned, with the innermost planet being most strongly affected.
The star would only need to be tilted by a few
degrees to explain the observed values of $\Delta I$.
It is not clear whether this scenario would result in
an association between higher mutual inclination and larger period ratios,
as we have observed.

In the ``secular chaos'' or ``high-$e$ migration'' scenario proposed by \citet{Petrovich},
the innermost planet of a multi-planet system 
is launched into a high-eccentricity orbit via chaotic secular interactions.
If the period is short and the
eccentricity becomes high enough ($\approx$0.8), tidal interactions with the host
star shrink the orbit.
Since eccentricity and inclination are excited together, this theory
predicts that the shortest-period planets should have larger mutual inclinations,
in qualitative agreement with our results.
A potential problem with this picture is that for systems in
mean-motion resonance (MMR), the dynamics
may be dominated by the resonance instead of secular
interactions. The sample of USP systems has a decent
fraction of systems that are in or near MMR
(5 out of 13 systems with at least 2 exterior
planets).

Another possibility is forced-eccentricity migration, in which
the interaction with outer companions (secular-forcing, or MMR) continually excites the eccentricity of the innermost planet.  This allows eccentricity tides to dissipate energy and shrink the orbit \citep[see, e.g.,][]{Hansen}. Since the planet's eccentricity never exceeds a few percent,
the inclinations are only excited to a few degrees, perhaps not enough
to be compatible with our results.

\acknowledgements
We thank Vincent Van Eylen, Daniel Fabrycky, Bonan Pu, Songhu Wang, and Cristobal Petrovich for helpful discussions. We are also grateful to the referee for a prompt report.

This work made use of the gaia-kepler.fun crossmatch database created by Megan Bedell. Work by F.D.\ and J.N.W.\ was supported by the Heising-Simons Foundation. Work by K.M.\ was performed in part under contract with the California Institute of Technology (Caltech)/Jet Propulsion Laboratory (JPL) funded by NASA through the Sagan Fellowship Program executed by the NASA Exoplanet Science Institute.

\startlongtable

\begin{deluxetable*}{ccccccccccccccc}
\tabletypesize{\scriptsize}
\tablecaption{Transit Modeling Results \label{tab:results}}
\tablehead{
\colhead{Planet} & \colhead{$\rho_\star$ (g cm$^{-3}$)} & \colhead{Source} &  \colhead{$P_1$ (days)} & \colhead{$P_2/P_1$} &  \colhead{$a/R_\star$} &   \colhead{$i (^{\circ})$}&  \colhead{$\Delta I  (^{\circ})$}&  \colhead{Cadence} &  \colhead{TTV}
}
\startdata
KOI-1843.03 & $ 5.115 ^{+ 0.155 }_{- 0.117 }$& \citet{2013ApJ...779..188M}+Gaia &$ 0.177 $&$ 23.72 $&$ 2.03 ^{+ 0.02 }_{- 0.02 } $&$ 87.17 ^{+ 2.10 }_{- 2.86 } $&$ 1.71 ^{+ 2.91 }_{- 1.22 } $& SC  & \\
EPIC-211305568b & $ 6.868 ^{+ 0.214 }_{- 0.186 }$& \citet{Dressing}+Gaia &$ 0.198 $&$ 58.39 $&$ 2.42 ^{+ 0.02 }_{- 0.02 } $&$ 77.71 ^{+ 7.35 }_{- 5.05 } $&$ 10.19 ^{+ 5.16 }_{- 7.18 } $& LC  & \\
K2-141b & $ 3.089 ^{+ 0.130 }_{- 0.110 }$& \citet{2018AaA...612A..95B}+Gaia &$ 0.280 $&$ 27.63 $&$ 2.33 ^{+ 0.03 }_{- 0.03 } $&$ 86.81 ^{+ 2.34 }_{- 3.41 } $&$ 2.08 ^{+ 2.17 }_{- 1.42 } $& LC  & \\
Kepler-42c & $ 56.740 ^{+ 15.410 }_{- 15.410 }$& \citet{2013ApJ...779..188M}+Gaia &$ 0.453 $&$ 2.68 $&$ 6.71 ^{+ 0.05 }_{- 0.11 } $&$ 89.14 ^{+ 0.63 }_{- 0.91 } $&$ 2.95 ^{+ 0.60 }_{- 0.77 } $& SC  & \\
K2-183b & $ 1.452 ^{+ 0.100 }_{- 0.087 }$& \citet{Mayo} &$ 0.469 $&$ 23.00 $&$ 2.55 ^{+ 0.05 }_{- 0.05 } $&$ 85.11 ^{+ 3.25 }_{- 4.03 } $&$ 4.19 ^{+ 3.93 }_{- 3.27 } $& LC  &  \\
K2-223b & $ 1.611 ^{+ 0.061 }_{- 0.072 }$& \citet{Mayo} &$ 0.506 $&$ 9.03 $&$ 2.79 ^{+ 0.04 }_{- 0.04 } $&$ 73.80 ^{+ 1.93 }_{- 1.15 } $&$ 14.72 ^{+ 1.50 }_{- 2.11 } $& LC  &  \\
Kepler-990c & $ 1.460 ^{+ 0.409 }_{- 0.343 }$& CKS+Gaia &$ 0.538 $&$ 18.42 $&$ 2.79 ^{+ 0.12 }_{- 0.13 } $&$ 84.86 ^{+ 3.67 }_{- 3.74 } $&$ 3.82 ^{+ 3.89 }_{- 2.96 } $& LC  & \\
K2-106b & $ 1.426 ^{+ 0.094 }_{- 0.086 }$& \citet{Guenther}+Gaia &$ 0.571 $&$ 23.38 $&$ 2.91 ^{+ 0.06 }_{- 0.06 } $&$ 82.65 ^{+ 5.03 }_{- 3.94 } $&$ 6.31 ^{+ 4.02 }_{- 4.82 } $& LC  & \\
K2-229b & $ 2.451 ^{+ 0.149 }_{- 0.147 }$& \citet{Santerne}+Gaia &$ 0.584 $&$ 14.25 $&$ 3.53 ^{+ 0.06 }_{- 0.07 } $&$ 87.01 ^{+ 2.14 }_{- 2.54 } $&$ 1.84 ^{+ 2.26 }_{- 1.28 } $& LC  & \\
KOI-787.03 & $ 1.104 ^{+ 0.154 }_{- 0.140 }$& \citet{Everett}+Gaia &$ 0.589 $&$ 7.52 $&$ 2.75 ^{+ 0.10 }_{- 0.11 } $&$ 80.86 ^{+ 5.82 }_{- 3.99 } $&$ 7.55 ^{+ 4.07 }_{- 5.61 } $& LC  & \\
KOI-2250.02 & $ 2.506 ^{+ 0.155 }_{- 0.178 }$& CKS+Gaia &$ 0.626 $&$ 4.69 $&$ 3.74 ^{+ 0.08 }_{- 0.08 } $&$ 84.51 ^{+ 3.14 }_{- 2.30 } $&$ 2.27 ^{+ 1.46 }_{- 1.49 } $& LC  & \\
Kepler-607b & $ 2.047 ^{+ 0.152 }_{- 0.108 }$& CKS+Gaia &$ 0.638 $&$ 62.18 $&$ 3.53 ^{+ 0.07 }_{- 0.07 } $&$ 83.68 ^{+ 3.98 }_{- 3.09 } $&$ 5.89 ^{+ 3.06 }_{- 3.96 } $& LC  & \\
EPIC-248435473b & $ 3.166 ^{+ 0.143 }_{- 0.105 }$& \citet{Rodriguez2018}+Gaia &$ 0.658 $&$ 9.27 $&$ 4.17 ^{+ 0.05 }_{- 0.05 } $&$ 76.40 ^{+ 0.26 }_{- 0.27 } $&$ 12.67 ^{+ 0.68 }_{- 0.75 } $& LC  & d, e \\
Kepler-1340b & $ 1.421 ^{+ 0.116 }_{- 0.134 }$& CKS+Gaia &$ 0.665 $&$ 7.62 $&$ 3.14 ^{+ 0.08 }_{- 0.07 } $&$ 87.62 ^{+ 1.73 }_{- 2.26 } $&$ 1.47 ^{+ 1.60 }_{- 1.07 } $& LC  & \\
KOI-191.03 & $ 1.535 ^{+ 0.179 }_{- 0.142 }$& CKS+Gaia &$ 0.709 $&$ 3.41 $&$ 3.49 ^{+ 0.07 }_{- 0.07 } $&$ 85.73 ^{+ 2.09 }_{- 1.41 } $&$ 1.39 ^{+ 1.13 }_{- 0.90 } $& SC  & \\
Kepler-32f & $ 4.577 ^{+ 0.164 }_{- 0.130 }$& \citet{2013ApJ...779..188M}+Gaia &$ 0.743 $&$ 3.90 $&$ 5.11 ^{+ 0.05 }_{- 0.05 } $&$ 87.68 ^{+ 1.38 }_{- 1.46 } $&$ 1.64 ^{+ 1.33 }_{- 1.16 } $& SC  & b, c, e \\
KOI-2248.03 & $ 2.343 ^{+ 0.365 }_{- 0.352 }$& \citet{Everett}+Gaia &$ 0.762 $&$ 3.47 $&$ 4.15 ^{+ 0.16 }_{- 0.19 } $&$ 85.58 ^{+ 2.85 }_{- 2.42 } $&$ 3.39 ^{+ 4.45 }_{- 2.32 } $& LC  & \\
Kepler-1067b & $ 1.792 ^{+ 0.098 }_{- 0.127 }$& CKS+Gaia &$ 0.762 $&$ 7.12 $&$ 3.80 ^{+ 0.09 }_{- 0.07 } $&$ 81.58 ^{+ 2.69 }_{- 1.73 } $&$ 6.75 ^{+ 1.67 }_{- 2.60 } $& LC  & \\
KOI-1360.03 & $ 2.879 ^{+ 0.131 }_{- 0.183 }$& CKS+Gaia &$ 0.764 $&$ 19.09 $&$ 4.46 ^{+ 0.08 }_{- 0.07 } $&$ 83.83 ^{+ 3.35 }_{- 2.34 } $&$ 5.23 ^{+ 2.37 }_{- 3.28 } $& LC  & \\
KOI-2393.02 & $ 2.809 ^{+ 0.148 }_{- 0.169 }$& CKS+Gaia &$ 0.767 $&$ 6.00 $&$ 4.44 ^{+ 0.08 }_{- 0.08 } $&$ 83.34 ^{+ 1.76 }_{- 1.11 } $&$ 5.59 ^{+ 1.30 }_{- 1.86 } $& LC  & \\
K2-187b & $ 1.889 ^{+ 0.113 }_{- 0.123 }$& \citet{Mayo} &$ 0.774 $&$ 3.71 $&$ 3.91 ^{+ 0.08 }_{- 0.08 } $&$ 82.61 ^{+ 2.48 }_{- 1.55 } $&$ 5.42 ^{+ 1.78 }_{- 2.60 } $& LC  &  \\
KOI-1239.01 & $ 1.628 ^{+ 0.183 }_{- 0.196 }$& CKS+Gaia &$ 0.783 $&$ 4.05 $&$ 3.70 ^{+ 0.09 }_{- 0.11 } $&$ 87.68 ^{+ 1.69 }_{- 2.18 } $&$ 1.42 ^{+ 1.36 }_{- 0.98 } $& LC  & \\
WASP-47e & $ 0.980 ^{+ 0.083 }_{- 0.079 }$& \citet{Becker}+Gaia &$ 0.790 $&$ 5.27 $&$ 3.20 ^{+ 0.04 }_{- 0.03 } $&$ 84.76 ^{+ 0.98 }_{- 0.78 } $&$ 4.00 ^{+ 0.64 }_{- 0.84 } $& SC  & b, c \\
Kepler-10b & $ 1.068 ^{+ 0.001 }_{- 0.007 }$& Astreoseismology &$ 0.837 $&$ 54.08 $&$ 3.406 ^{+ 0.005 }_{- 0.005 } $&$ 84.02 ^{+ 0.12 }_{- 0.11 } $&$ 5.82 ^{+ 0.16 }_{- 0.17 } $& SC  &  c\\
Kepler-10b & $ 1.010 ^{+ 0.077 }_{- 0.063 }$& CKS+Gaia &$ 0.838 $&$ 54.08 $&$ 3.32 ^{+ 0.07 }_{- 0.06 } $&$ 82.78 ^{+ 1.03 }_{- 0.85 } $&$ 7.03 ^{+ 0.84 }_{- 1.02 } $& SC  & c \\
KOI-1499.02 & $ 2.060 ^{+ 0.179 }_{- 0.185 }$& KIC\footnote{\citet{Mathur}}+Gaia &$ 0.841 $&$ 7.39 $&$ 4.25 ^{+ 0.12 }_{- 0.13 } $&$ 84.70 ^{+ 2.95 }_{- 2.25 } $&$ 3.91 ^{+ 2.33 }_{- 2.69 } $& LC  & \\
Kepler-732c & $ 5.421 ^{+ 0.150 }_{- 0.122 }$& \citet{2013ApJ...779..188M}+Gaia &$ 0.893 $&$ 10.60 $&$ 6.12 ^{+ 0.05 }_{- 0.05 } $&$ 87.92 ^{+ 0.64 }_{- 0.48 } $&$ 1.71 ^{+ 0.55 }_{- 0.67 } $& SC  & \\
Kepler-653c & $ 0.807 ^{+ 0.062 }_{- 0.059 }$& CKS+Gaia &$ 0.900 $&$ 16.34 $&$ 3.26 ^{+ 0.08 }_{- 0.07 } $&$ 75.99 ^{+ 1.68 }_{- 1.16 } $&$ 12.38 ^{+ 1.52 }_{- 1.72 } $& LC  & \\
EPIC-206024342b & $ 1.994 ^{+ 0.131 }_{- 0.167 }$& \citet{2018AJ....155...84W}+Gaia &$ 0.912 $&$ 16.06 $&$ 4.42 ^{+ 0.11 }_{- 0.10 } $&$ 84.60 ^{+ 3.62 }_{- 3.16 } $&$ 3.95 ^{+ 3.05 }_{- 3.06 } $& LC  & \\
HD-3167b & $ 1.916 ^{+ 0.114 }_{- 0.107 }$& \citet{VanderburgHD3167}+Gaia &$ 0.960 $&$ 31.10 $&$ 4.53 ^{+ 0.08 }_{- 0.08 } $&$ 87.38 ^{+ 1.69 }_{- 1.50 } $&$ 2.04 ^{+ 1.46 }_{- 1.54 } $& LC  & \\
Kepler-1322b & $ 1.437 ^{+ 0.228 }_{- 0.240 }$& KIC+Gaia &$ 0.963 $&$ 6.71 $&$ 4.15 ^{+ 0.19 }_{- 0.19 } $&$ 85.59 ^{+ 2.60 }_{- 2.35 } $&$ 2.10 ^{+ 1.89 }_{- 1.50 } $& LC  & \\
KOI-3145.02 & $ 3.129 ^{+ 0.189 }_{- 0.183 }$& KIC+Gaia &$ 0.977 $&$ 4.64 $&$ 5.41 ^{+ 0.12 }_{- 0.10 } $&$ 86.64 ^{+ 2.23 }_{- 2.24 } $&$ 1.80 ^{+ 2.12 }_{- 1.21 } $& LC  & \\
Kepler-80f & $ 3.026 ^{+ 0.107 }_{- 0.136 }$& CKS+Gaia &$ 0.987 $&$ 3.11 $&$ 5.35 ^{+ 0.05 }_{- 0.06 } $&$ 88.85 ^{+ 0.89 }_{- 1.00 } $&$ 0.89 ^{+ 0.90 }_{- 0.65 } $& SC  & b, c \\
Kepler-755b & $ 2.031 ^{+ 0.149 }_{- 0.096 }$& CKS+Gaia &$ 1.269 $&$ 2.25 $&$ 5.57 ^{+ 0.11 }_{- 0.11 } $&$ 84.12 ^{+ 0.41 }_{- 0.36 } $&$ 1.58 ^{+ 2.67 }_{- 1.07 } $& SC  & \\
Kepler-198d & $ 1.695 ^{+ 0.146 }_{- 0.145 }$& KIC+Gaia &$ 1.312 $&$ 13.56 $&$ 5.34 ^{+ 0.16 }_{- 0.12 } $&$ 87.15 ^{+ 2.04 }_{- 1.45 } $&$ 2.43 ^{+ 1.41 }_{- 1.91 } $& SC  & \\
Kepler-207b & $ 0.334 ^{+ 0.027 }_{- 0.025 }$& CKS+Gaia &$ 1.612 $&$ 1.91 $&$ 3.60 ^{+ 0.10 }_{- 0.10 } $&$ 81.77 ^{+ 1.55 }_{- 1.23 } $&$ 5.93 ^{+ 1.91 }_{- 2.33 } $& LC  & \\
Kepler-342e & $ 0.661 ^{+ 0.056 }_{- 0.054 }$& CKS+Gaia &$ 1.644 $&$ 9.23 $&$ 4.46 ^{+ 0.09 }_{- 0.11 } $&$ 88.67 ^{+ 1.00 }_{- 1.51 } $&$ 0.85 ^{+ 1.19 }_{- 0.58 } $& LC  & c, d \\
Kepler-322b & $ 2.222 ^{+ 0.119 }_{- 0.151 }$& CKS+Gaia &$ 1.654 $&$ 2.62 $&$ 6.86 ^{+ 0.13 }_{- 0.14 } $&$ 86.85 ^{+ 1.32 }_{- 0.84 } $&$ 2.12 ^{+ 0.92 }_{- 1.21 } $& LC  & \\
Kepler-323b & $ 1.135 ^{+ 0.109 }_{- 0.102 }$& KIC+Gaia &$ 1.678 $&$ 2.12 $&$ 5.49 ^{+ 0.08 }_{- 0.12 } $&$ 88.51 ^{+ 1.01 }_{- 1.37 } $&$ 0.82 ^{+ 1.01 }_{- 0.55 } $& LC  & \\
Kepler-969c & $ 2.375 ^{+ 0.113 }_{- 0.146 }$& CKS+Gaia &$ 1.683 $&$ 20.31 $&$ 7.08 ^{+ 0.11 }_{- 0.11 } $&$ 88.68 ^{+ 0.88 }_{- 0.96 } $&$ 0.82 ^{+ 0.96 }_{- 0.60 } $& LC  & \\
Kepler-1047c & $ 0.371 ^{+ 0.037 }_{- 0.030 }$& CKS+Gaia &$ 1.721 $&$ 1.85 $&$ 3.87 ^{+ 0.12 }_{- 0.10 } $&$ 78.25 ^{+ 1.78 }_{- 1.20 } $&$ 6.30 ^{+ 3.55 }_{- 2.84 } $& LC  & \\
Kepler-312b & $ 0.371 ^{+ 0.031 }_{- 0.030 }$& CKS+Gaia &$ 1.772 $&$ 11.14 $&$ 3.95 ^{+ 0.11 }_{- 0.10 } $&$ 79.32 ^{+ 0.62 }_{- 0.64 } $&$ 9.12 ^{+ 1.19 }_{- 1.38 } $& LC  & c \\
Kepler-524c & $ 0.410 ^{+ 0.037 }_{- 0.031 }$& CKS+Gaia &$ 1.889 $&$ 4.22 $&$ 4.26 ^{+ 0.10 }_{- 0.10 } $&$ 87.24 ^{+ 1.91 }_{- 1.97 } $&$ 1.65 ^{+ 1.82 }_{- 1.16 } $& LC  & b \\
Kepler-1371c & $ 1.440 ^{+ 0.086 }_{- 0.066 }$& CKS+Gaia &$ 2.005 $&$ 1.45 $&$ 6.73 ^{+ 0.11 }_{- 0.12 } $&$ 85.91 ^{+ 2.20 }_{- 1.08 } $&$ 1.59 ^{+ 1.65 }_{- 1.08 } $& LC  & \\
Kepler-142b & $ 1.012 ^{+ 0.087 }_{- 0.074 }$& CKS+Gaia &$ 2.024 $&$ 2.35 $&$ 6.06 ^{+ 0.10 }_{- 0.15 } $&$ 88.02 ^{+ 1.21 }_{- 1.02 } $&$ 0.89 ^{+ 0.81 }_{- 0.62 } $& SC  & \\
Kepler-326b & $ 1.068 ^{+ 0.073 }_{- 0.078 }$& CKS+Gaia &$ 2.248 $&$ 2.04 $&$ 6.62 ^{+ 0.16 }_{- 0.15 } $&$ 84.59 ^{+ 0.90 }_{- 0.83 } $&$ 3.59 ^{+ 1.36 }_{- 1.51 } $& LC  & \\
Kepler-406b & $ 1.133 ^{+ 0.105 }_{- 0.090 }$& CKS+Gaia &$ 2.426 $&$ 1.91 $&$ 6.97 ^{+ 0.13 }_{- 0.15 } $&$ 89.07 ^{+ 0.62 }_{- 0.92 } $&$ 1.02 ^{+ 0.97 }_{- 0.77 } $& LC  & \\
Kepler-314b & $ 1.845 ^{+ 0.137 }_{- 0.134 }$& CKS+Gaia &$ 2.461 $&$ 2.42 $&$ 8.35 ^{+ 0.19 }_{- 0.18 } $&$ 88.65 ^{+ 0.88 }_{- 1.05 } $&$ 0.79 ^{+ 0.73 }_{- 0.52 } $& LC  & \\
Kepler-213b & $ 1.110 ^{+ 0.098 }_{- 0.096 }$& CKS+Gaia &$ 2.462 $&$ 1.96 $&$ 7.10 ^{+ 0.22 }_{- 0.18 } $&$ 83.99 ^{+ 0.37 }_{- 0.32 } $&$ 0.94 ^{+ 0.37 }_{- 0.37 } $& SC  & \\
Kepler-1311c & $ 0.298 ^{+ 0.022 }_{- 0.018 }$& CKS+Gaia &$ 2.536 $&$ 4.41 $&$ 4.72 ^{+ 0.08 }_{- 0.08 } $&$ 87.22 ^{+ 1.65 }_{- 1.47 } $&$ 1.53 ^{+ 1.60 }_{- 1.03 } $& LC  & \\
Kepler-1398b & $ 0.611 ^{+ 0.048 }_{- 0.044 }$& CKS+Gaia &$ 2.788 $&$ 1.48 $&$ 6.27 ^{+ 0.13 }_{- 0.15 } $&$ 88.45 ^{+ 1.03 }_{- 1.22 } $&$ 1.45 ^{+ 1.29 }_{- 0.98 } $& LC  & \\
Kepler-221b & $ 2.417 ^{+ 0.132 }_{- 0.159 }$& CKS+Gaia &$ 2.796 $&$ 2.04 $&$ 10.00 ^{+ 0.22 }_{- 0.19 } $&$ 88.18 ^{+ 0.51 }_{- 0.32 } $&$ 0.98 ^{+ 0.51 }_{- 0.51 } $& SC  & e \\
Kepler-1542c & $ 0.837 ^{+ 0.067 }_{- 0.051 }$& CKS+Gaia &$ 2.892 $&$ 1.37 $&$ 7.18 ^{+ 0.15 }_{- 0.15 } $&$ 85.49 ^{+ 0.72 }_{- 0.52 } $&$ 1.77 ^{+ 1.88 }_{- 1.34 } $& LC  & \\
Kepler-411b & $ 2.630 ^{+ 0.132 }_{- 0.139 }$& CKS+Gaia &$ 3.005 $&$ 2.61 $&$ 10.78 ^{+ 0.17 }_{- 0.13 } $&$ 87.71 ^{+ 0.19 }_{- 0.16 } $&$ 1.23 ^{+ 0.14 }_{- 0.15 } $& SC  & c \\
Kepler-1271b & $ 0.951 ^{+ 0.105 }_{- 0.090 }$& KIC+Gaia &$ 3.026 $&$ 1.79 $&$ 7.74 ^{+ 0.24 }_{- 0.23 } $&$ 88.10 ^{+ 1.22 }_{- 1.03 } $&$ 1.04 ^{+ 1.10 }_{- 0.72 } $& LC  & \\
Kepler-141b & $ 2.311 ^{+ 0.153 }_{- 0.106 }$& CKS+Gaia &$ 3.108 $&$ 2.26 $&$ 10.49 ^{+ 0.17 }_{- 0.18 } $&$ 89.24 ^{+ 0.50 }_{- 0.70 } $&$ 0.52 ^{+ 0.53 }_{- 0.36 } $& LC  & \\
Kepler-203b & $ 1.167 ^{+ 0.120 }_{- 0.107 }$& CKS+Gaia &$ 3.163 $&$ 1.70 $&$ 8.53 ^{+ 0.24 }_{- 0.25 } $&$ 84.92 ^{+ 0.28 }_{- 0.28 } $&$ 3.39 ^{+ 1.18 }_{- 1.30 } $& LC  & \\
Kepler-107b & $ 0.581 ^{+ 0.049 }_{- 0.049 }$& Astreoseismology &$ 3.180 $&$ 1.54 $&$ 6.71 ^{+ 0.06 }_{- 0.11 } $&$ 88.97 ^{+ 0.65 }_{- 0.86 } $&$ 1.62 ^{+ 0.80 }_{- 0.94 } $& SC  & \\
Kepler-107b & $ 0.519 ^{+ 0.038 }_{- 0.045 }$& CKS+Gaia &$ 3.180 $&$ 1.54 $&$ 6.54 ^{+ 0.15 }_{- 0.14 } $&$ 87.74 ^{+ 1.09 }_{- 0.69 } $&$ 0.99 ^{+ 0.89 }_{- 0.66 } $& SC  & \\
Kepler-140b & $ 0.904 ^{+ 0.080 }_{- 0.067 }$& CKS+Gaia &$ 3.254 $&$ 28.07 $&$ 8.01 ^{+ 0.23 }_{- 0.20 } $&$ 87.45 ^{+ 1.08 }_{- 0.55 } $&$ 2.31 ^{+ 0.57 }_{- 1.06 } $& LC  & \\
Kepler-337b & $ 0.278 ^{+ 0.023 }_{- 0.019 }$& CKS+Gaia &$ 3.293 $&$ 2.94 $&$ 5.45 ^{+ 0.13 }_{- 0.12 } $&$ 86.51 ^{+ 1.57 }_{- 0.87 } $&$ 2.16 ^{+ 1.27 }_{- 1.31 } $& SC  & \\
Kepler-111b & $ 1.002 ^{+ 0.085 }_{- 0.078 }$& CKS+Gaia &$ 3.342 $&$ 67.26 $&$ 8.48 ^{+ 0.18 }_{- 0.22 } $&$ 88.40 ^{+ 0.77 }_{- 0.67 } $&$ 1.32 ^{+ 0.65 }_{- 0.75 } $& SC  & \\
Kepler-101b & $ 0.311 ^{+ 0.027 }_{- 0.022 }$& CKS+Gaia &$ 3.488 $&$ 1.73 $&$ 5.87 ^{+ 0.17 }_{- 0.15 } $&$ 84.22 ^{+ 0.73 }_{- 0.46 } $&$ 2.19 ^{+ 2.23 }_{- 1.31 } $& SC  & b \\
Kepler-18b & $ 1.612 ^{+ 0.143 }_{- 0.130 }$& CKS+Gaia &$ 3.505 $&$ 2.18 $&$ 10.14 ^{+ 0.24 }_{- 0.20 } $&$ 86.32 ^{+ 0.21 }_{- 0.19 } $&$ 1.89 ^{+ 0.12 }_{- 0.14 } $& SC  & c, d \\
Kepler-363b & $ 0.396 ^{+ 0.034 }_{- 0.033 }$& CKS+Gaia &$ 3.615 $&$ 2.09 $&$ 6.49 ^{+ 0.17 }_{- 0.18 } $&$ 84.02 ^{+ 0.64 }_{- 0.54 } $&$ 4.09 ^{+ 1.36 }_{- 1.06 } $& LC  & \\
Kepler-218b & $ 1.116 ^{+ 0.095 }_{- 0.072 }$& CKS+Gaia &$ 3.619 $&$ 12.35 $&$ 9.16 ^{+ 0.19 }_{- 0.19 } $&$ 88.77 ^{+ 0.86 }_{- 0.81 } $&$ 0.80 ^{+ 0.81 }_{- 0.58 } $& LC  & \\
Kepler-20b & $ 1.756 ^{+ 0.155 }_{- 0.132 }$& CKS+Gaia &$ 3.696 $&$ 1.65 $&$ 11.00 ^{+ 0.15 }_{- 0.17 } $&$ 88.19 ^{+ 0.23 }_{- 0.22 } $&$ 0.20 ^{+ 0.35 }_{- 0.16 } $& SC  & b, c \\
Kepler-466c & $ 1.505 ^{+ 0.241 }_{- 0.259 }$& CKS+Gaia &$ 3.709 $&$ 13.77 $&$ 10.50 ^{+ 0.32 }_{- 0.37 } $&$ 88.76 ^{+ 0.79 }_{- 0.78 } $&$ 1.02 ^{+ 0.81 }_{- 0.71 } $& SC  & \\
Kepler-89b & $ 0.596 ^{+ 0.055 }_{- 0.047 }$& CKS+Gaia &$ 3.743 $&$ 2.78 $&$ 7.70 ^{+ 0.13 }_{- 0.10 } $&$ 88.09 ^{+ 0.62 }_{- 0.39 } $&$ 0.28 ^{+ 0.38 }_{- 0.18 } $& SC  & c, d, e \\
Kepler-217d & $ 0.289 ^{+ 0.023 }_{- 0.023 }$& CKS+Gaia &$ 3.887 $&$ 1.38 $&$ 6.13 ^{+ 0.16 }_{- 0.16 } $&$ 84.18 ^{+ 0.64 }_{- 0.52 } $&$ 3.53 ^{+ 1.52 }_{- 1.86 } $& LC  & \\
Kepler-380b & $ 0.757 ^{+ 0.074 }_{- 0.060 }$& CKS+Gaia &$ 3.931 $&$ 1.94 $&$ 8.54 ^{+ 0.24 }_{- 0.24 } $&$ 86.31 ^{+ 0.80 }_{- 0.53 } $&$ 1.10 ^{+ 1.54 }_{- 0.78 } $& LC  & \\
Kepler-402b & $ 1.068 ^{+ 0.099 }_{- 0.092 }$& CKS+Gaia &$ 4.029 $&$ 1.52 $&$ 9.74 ^{+ 0.29 }_{- 0.26 } $&$ 88.06 ^{+ 1.03 }_{- 0.67 } $&$ 0.77 ^{+ 0.74 }_{- 0.50 } $& LC  & \\
Kepler-202b & $ 3.014 ^{+ 0.115 }_{- 0.146 }$& CKS+Gaia &$ 4.069 $&$ 4.00 $&$ 13.82 ^{+ 0.20 }_{- 0.20 } $&$ 88.06 ^{+ 0.16 }_{- 0.14 } $&$ 0.91 ^{+ 0.16 }_{- 0.14 } $& SC  & \\
Kepler-625c & $ 0.457 ^{+ 0.047 }_{- 0.035 }$& KIC+Gaia &$ 4.165 $&$ 1.86 $&$ 7.49 ^{+ 0.23 }_{- 0.21 } $&$ 85.06 ^{+ 0.36 }_{- 0.37 } $&$ 3.34 ^{+ 1.08 }_{- 1.25 } $& LC  & \\
Kepler-208b & $ 0.745 ^{+ 0.068 }_{- 0.060 }$& CKS+Gaia &$ 4.229 $&$ 1.77 $&$ 8.90 ^{+ 0.26 }_{- 0.22 } $&$ 87.30 ^{+ 0.87 }_{- 0.50 } $&$ 0.68 ^{+ 0.71 }_{- 0.43 } $& LC  & \\
Kepler-783b & $ 1.939 ^{+ 0.165 }_{- 0.142 }$& CKS+Gaia &$ 4.293 $&$ 1.64 $&$ 12.36 ^{+ 0.34 }_{- 0.30 } $&$ 88.47 ^{+ 0.98 }_{- 0.81 } $&$ 1.26 ^{+ 1.39 }_{- 0.88 } $& LC  & \\
Kepler-219b & $ 0.840 ^{+ 0.075 }_{- 0.069 }$& CKS+Gaia &$ 4.585 $&$ 4.95 $&$ 9.84 ^{+ 0.28 }_{- 0.27 } $&$ 87.46 ^{+ 0.48 }_{- 0.37 } $&$ 1.85 ^{+ 0.36 }_{- 0.43 } $& SC  & b \\
Kepler-356b & $ 0.562 ^{+ 0.049 }_{- 0.046 }$& KIC+Gaia &$ 4.613 $&$ 2.84 $&$ 8.59 ^{+ 0.25 }_{- 0.22 } $&$ 85.50 ^{+ 0.34 }_{- 0.29 } $&$ 1.80 ^{+ 0.30 }_{- 0.32 } $& LC  & \\
Kepler-1365c & $ 0.360 ^{+ 0.032 }_{- 0.027 }$& CKS+Gaia &$ 4.775 $&$ 1.61 $&$ 7.57 ^{+ 0.21 }_{- 0.21 } $&$ 85.32 ^{+ 0.70 }_{- 0.53 } $&$ 2.09 ^{+ 1.54 }_{- 0.95 } $& LC  & \\
Kepler-321b & $ 1.441 ^{+ 0.120 }_{- 0.120 }$& CKS+Gaia &$ 4.915 $&$ 2.66 $&$ 12.30 ^{+ 0.36 }_{- 0.30 } $&$ 87.67 ^{+ 0.28 }_{- 0.24 } $&$ 0.25 ^{+ 0.21 }_{- 0.16 } $& LC  & b \\
Kepler-376b & $ 0.500 ^{+ 0.047 }_{- 0.040 }$& KIC+Gaia &$ 4.920 $&$ 2.88 $&$ 8.70 ^{+ 0.23 }_{- 0.23 } $&$ 87.79 ^{+ 0.99 }_{- 0.70 } $&$ 0.80 ^{+ 0.97 }_{- 0.60 } $& LC  & \\
Kepler-634b & $ 0.472 ^{+ 0.045 }_{- 0.044 }$& KIC+Gaia &$ 5.169 $&$ 1.57 $&$ 8.69 ^{+ 0.17 }_{- 0.20 } $&$ 88.93 ^{+ 0.72 }_{- 0.82 } $&$ 1.06 ^{+ 0.96 }_{- 0.69 } $& LC  & \\
Kepler-392b & $ 0.881 ^{+ 0.069 }_{- 0.066 }$& CKS+Gaia &$ 5.342 $&$ 1.47 $&$ 10.97 ^{+ 0.23 }_{- 0.25 } $&$ 88.97 ^{+ 0.72 }_{- 0.76 } $&$ 0.78 ^{+ 0.77 }_{- 0.52 } $& LC  & \\
Kepler-526b & $ 0.478 ^{+ 0.038 }_{- 0.038 }$& CKS+Gaia &$ 5.459 $&$ 1.26 $&$ 9.11 ^{+ 0.24 }_{- 0.24 } $&$ 86.89 ^{+ 0.49 }_{- 0.36 } $&$ 1.16 ^{+ 1.26 }_{- 0.84 } $& LC  & \\
Kepler-197b & $ 0.907 ^{+ 0.052 }_{- 0.052 }$& Astreoseismology &$ 5.599 $&$ 1.85 $&$ 11.46 ^{+ 0.17 }_{- 0.20 } $&$ 89.06 ^{+ 0.59 }_{- 0.52 } $&$ 0.42 ^{+ 0.53 }_{- 0.30 } $& SC  & \\
Kepler-197b & $ 0.815 ^{+ 0.067 }_{- 0.059 }$& CKS+Gaia &$ 5.599 $&$ 1.85 $&$ 11.13 ^{+ 0.29 }_{- 0.24 } $&$ 88.57 ^{+ 0.73 }_{- 0.52 } $&$ 0.35 ^{+ 0.45 }_{- 0.24 } $& SC  & \\
Kepler-381b & $ 0.482 ^{+ 0.045 }_{- 0.045 }$& CKS+Gaia &$ 5.629 $&$ 1.47 $&$ 9.31 ^{+ 0.31 }_{- 0.27 } $&$ 86.62 ^{+ 0.52 }_{- 0.41 } $&$ 2.77 ^{+ 1.44 }_{- 1.81 } $& LC  & \\
Kepler-33b & $ 0.353 ^{+ 0.031 }_{- 0.027 }$& CKS+Gaia &$ 5.668 $&$ 2.32 $&$ 8.38 ^{+ 0.16 }_{- 0.20 } $&$ 88.35 ^{+ 1.24 }_{- 0.98 } $&$ 0.76 ^{+ 1.02 }_{- 0.53 } $& SC  & d, e, f \\
Kepler-116b & $ 0.575 ^{+ 0.054 }_{- 0.048 }$& CKS+Gaia &$ 5.969 $&$ 2.19 $&$ 10.32 ^{+ 0.29 }_{- 0.29 } $&$ 86.63 ^{+ 0.27 }_{- 0.26 } $&$ 0.71 ^{+ 1.71 }_{- 0.59 } $& SC  & \\
Kepler-135b & $ 0.630 ^{+ 0.056 }_{- 0.054 }$& KIC+Gaia &$ 6.003 $&$ 1.91 $&$ 10.68 ^{+ 0.30 }_{- 0.29 } $&$ 87.69 ^{+ 0.71 }_{- 0.46 } $&$ 1.01 ^{+ 0.93 }_{- 0.71 } $& SC  & \\
Kepler-132b & $ 1.237 ^{+ 0.253 }_{- 0.212 }$& CKS+Gaia &$ 6.178 $&$ 2.92 $&$ 13.59 ^{+ 0.72 }_{- 0.62 } $&$ 88.49 ^{+ 0.88 }_{- 0.59 } $&$ 0.69 ^{+ 0.69 }_{- 0.51 } $& LC  & d \\
Kepler-1581b & $ 0.572 ^{+ 0.047 }_{- 0.045 }$& KIC+Gaia &$ 6.284 $&$ 1.45 $&$ 10.63 ^{+ 0.28 }_{- 0.25 } $&$ 87.99 ^{+ 1.00 }_{- 0.68 } $&$ 0.88 ^{+ 0.96 }_{- 0.60 } $& LC  & \\
Kepler-335b & $ 0.281 ^{+ 0.024 }_{- 0.022 }$& CKS+Gaia &$ 6.562 $&$ 10.34 $&$ 8.61 ^{+ 0.21 }_{- 0.21 } $&$ 84.97 ^{+ 0.28 }_{- 0.23 } $&$ 4.54 ^{+ 0.34 }_{- 0.31 } $& LC  & c \\
Kepler-431b & $ 0.410 ^{+ 0.032 }_{- 0.028 }$& CKS+Gaia &$ 6.802 $&$ 1.28 $&$ 10.02 ^{+ 0.24 }_{- 0.21 } $&$ 87.41 ^{+ 0.84 }_{- 0.59 } $&$ 0.90 ^{+ 1.13 }_{- 0.65 } $& LC  & \\
Kepler-100b & $ 0.454 ^{+ 0.004 }_{- 0.006 }$& Astreoseismology &$ 6.887 $&$ 1.86 $&$ 10.43 ^{+ 0.04 }_{- 0.04 } $&$ 87.32 ^{+ 0.06 }_{- 0.06 } $&$ 1.39 ^{+ 0.54 }_{- 0.38 } $& SC  & c \\
Kepler-100b & $ 0.428 ^{+ 0.033 }_{-0.027  }$& CKS+Gaia &$ 6.887 $&$ 1.86 $&$ 10.24 ^{+ 0.23 }_{- 0.19 } $&$ 87.12 ^{+ 0.26 }_{- 0.21 } $&$ 1.66 ^{+ 0.46 }_{- 0.36 } $& SC  & c \\
Kepler-403b & $ 0.317 ^{+ 0.025 }_{- 0.022 }$& CKS+Gaia &$ 7.031 $&$ 1.94 $&$ 9.25 ^{+ 0.17 }_{- 0.17 } $&$ 89.22 ^{+ 0.58 }_{- 0.70 } $&$ 0.56 ^{+ 0.63 }_{- 0.40 } $& LC  & \\
Kepler-60b & $ 0.491 ^{+ 0.044 }_{- 0.040 }$& CKS+Gaia &$ 7.133 $&$ 1.25 $&$ 11.00 ^{+ 0.30 }_{- 0.29 } $&$ 88.15 ^{+ 1.06 }_{- 0.66 } $&$ 0.73 ^{+ 1.08 }_{- 0.53 } $& LC  & b, c, d \\
Kepler-853b & $ 0.540 ^{+ 0.045 }_{- 0.043 }$& CKS+Gaia &$ 7.169 $&$ 7.58 $&$ 11.38 ^{+ 0.29 }_{- 0.30 } $&$ 88.43 ^{+ 0.82 }_{- 0.58 } $&$ 1.08 ^{+ 0.66 }_{- 0.72 } $& LC  & \\
Kepler-450d & $ 0.478 ^{+ 0.064 }_{- 0.064 }$& Astreoseismology &$ 7.514 $&$ 2.05 $&$ 11.29 ^{+ 0.32 }_{- 0.30 } $&$ 88.55 ^{+ 0.73 }_{- 0.50 } $&$ 0.39 ^{+ 0.40 }_{- 0.26 } $& SC  & b, c \\
Kepler-450d & $ 0.438 ^{+ 0.029 }_{- 0.028 }$& CKS+Gaia &$ 7.514 $&$ 2.05 $&$ 11.06 ^{+ 0.20 }_{- 0.18 } $&$ 88.16 ^{+ 0.37 }_{- 0.29 } $&$ 0.99 ^{+ 0.38 }_{- 0.40 } $& SC  & b, c \\
Kepler-216b & $ 0.425 ^{+ 0.038 }_{- 0.034 }$& CKS+Gaia &$ 7.694 $&$ 2.26 $&$ 11.09 ^{+ 0.28 }_{- 0.30 } $&$ 88.68 ^{+ 0.71 }_{- 0.54 } $&$ 0.65 ^{+ 0.59 }_{- 0.47 } $& SC  & \\
Kepler-200b & $ 1.300 ^{+ 0.124 }_{- 0.109 }$& CKS+Gaia &$ 8.595 $&$ 1.19 $&$ 17.19 ^{+ 0.53 }_{- 0.52 } $&$ 88.34 ^{+ 0.39 }_{- 0.30 } $&$ 0.84 ^{+ 0.56 }_{- 0.41 } $& LC  & \\
Kepler-338e & $ 0.309 ^{+ 0.034 }_{- 0.034 }$& Astreoseismology &$ 9.342 $&$ 1.47 $&$ 11.28 ^{+ 0.31 }_{- 0.36 } $&$ 88.50 ^{+ 0.69 }_{- 0.52 } $&$ 0.74 ^{+ 0.68 }_{- 0.49 } $& SC  & \\
Kepler-338e & $ 0.293 ^{+ 0.024 }_{- 0.022 }$& CKS+Gaia &$ 9.342 $&$ 1.47 $&$ 11.07 ^{+ 0.29 }_{- 0.23 } $&$ 88.20 ^{+ 0.53 }_{- 0.37 } $&$ 1.00 ^{+ 0.58 }_{- 0.66 } $& SC  & \\
Kepler-804c & $ 1.168 ^{+ 0.109 }_{- 0.098 }$& CKS+Gaia &$ 9.652 $&$ 1.49 $&$ 17.67 ^{+ 0.42 }_{- 0.41 } $&$ 89.56 ^{+ 0.32 }_{- 0.39 } $&$ 0.36 ^{+ 0.40 }_{- 0.26 } $& LC  & \\
Kepler-36b & $ 0.361 ^{+ 0.024 }_{- 0.022 }$& CKS+Gaia &$ 13.850 $&$ 1.17 $&$ 15.41 ^{+ 0.13 }_{- 0.19 } $&$ 89.48 ^{+ 0.32 }_{- 0.30 } $&$ 0.29 ^{+ 0.28 }_{- 0.21 } $& SC  & b, c \\
Kepler-277b & $ 0.316 ^{+ 0.027 }_{- 0.025 }$& CKS+Gaia &$ 17.324 $&$ 1.91 $&$ 15.99 ^{+ 0.13 }_{- 0.18 } $&$ 89.75 ^{+ 0.17 }_{- 0.25 } $&$ 1.06 ^{+ 0.64 }_{- 0.80 } $& SC  & b \\
\enddata
\end{deluxetable*}

\begin{deluxetable*}{lcll}
\tablecaption{Model comparison with Hierarchical Bayesian Modeling \label{tab:HBM}}
\tablehead{
\colhead{Model for $\Delta I$} & \colhead{Bayesian Evidence log$(Z)$} & \colhead{Parameters}&  \colhead{Prior}
}
\startdata
1: $\Delta I \sim P(\sigma_0)$\footnote{PDF of a Rayleigh distribution} &  $-41.9$  &  $\sigma_0 =$ 0.0504$^{+0.0024}_{-0.0025}$ (RMS $\Delta I = 4.05^{+0.19}_{-0.20}~^{\circ})$ & $\sigma: $ log-uniform [-5,5] \\
\hline
2: $\Delta I \sim P(\sigma_0 (\frac{a}{R_\star})^m)$ &  $-8.3$&$\sigma_0=0.382^{+0.080}_{-0.063}$    &$\sigma_0: $ log-uniform [-5,5]\\
 & & $m=-1.28\pm 0.10$ & $m$: uniform [-5,5]\\
\hline
3: $\Delta I \sim P(\sigma_1)$ if $\frac{a}{R_\star}<\frac{a}{R_\star}^\prime$  & $-15.1$ & $\sigma_1=$ 0.0830$^{+0.0079}_{-0.0069}$ (RMS $\Delta I = 6.68^{+0.64}_{-0.55}~^{\circ})$  &$\sigma_1: $ log-uniform [-5,5] \\
$\Delta I \sim P(\sigma_2)$ if $\frac{a}{R_\star}>\frac{a}{R_\star}^\prime$ & &$\sigma_2=$0.0250$^{+ 0.0016}_{-0.0014}$ (RMS $\Delta I = 
2.01^{+0.13}_{-0.11}~^{\circ})$  &$\sigma_2: $ log-uniform [-5,5]\\
&& $\frac{a}{R_\star}^\prime=$ 4.65$^{+ 0.27}_{-0.10}$& $\frac{a}{R_\star}^\prime$: log-uniform [0,2]\\
\hline
4: $\Delta I \sim P(\sigma_1)$ if $\frac{P_2}{P_1}<\frac{P_2}{P_1}^\prime$  & $-27.1$ & $\sigma_1=$ 0.0311$^{+ 0.0021}_{-0.0022}$ (RMS $\Delta I = 2.50^{+0.17}_{-0.18}~^{\circ})$ &$\sigma_1: $ log-uniform [-5,5] \\
$\Delta I \sim P(\sigma_2)$ if $\frac{P_2}{P_1}>\frac{P_2}{P_1}^\prime$ &   & $\sigma_2=$ 0.0765$^{+0.0081}_{-0.0066}$ (RMS $\Delta I = 6.15^{+0.65}_{-0.53}~^{\circ})$ &$\sigma_2: $ log-uniform [-5,5]  \\
&&  $\frac{P_2}{P_1}^\prime=$  5.38$^{+0.56}_{-0.86}$& $\frac{P_2}{P_1}^\prime$: log-uniform [0,2]\\
\enddata
\end{deluxetable*}



\end{document}